\documentclass[11pt,a4paper]{article}
\usepackage{jcappub}

\newcommand{\bea}{\begin{eqnarray}}
\newcommand{\ena}{\end{eqnarray}}

\newcommand{\RR}{{\cal R}}

\newcommand{\PP}{{\cal P}}
\newcommand{\LL}{{\cal L}}

\renewcommand{\O}{\Omega}

\title{Reconstruction of the primordial power spectrum from CMB data}

\author[a,b]{Zong-Kuan Guo,}
\author[b]{Dominik J. Schwarz}
\author[a]{and Yuan-Zhong Zhang}

\affiliation[a]{Key Laboratory of Frontiers in Theoretical Physics, Institute of Theoretical Physics,\\
  Chinese Academy of Sciences, P.O. Box 2735, Beijing 100190, China}
\affiliation[b]{Fakult{\"a}t f{\"u}r Physik, Universit{\"a}t Bielefeld,
  Postfach 100131, 33501 Bielefeld, Germany}

\emailAdd{guozk@itp.ac.cn}
\emailAdd{dschwarz@physik.uni-bielefeld.de}
\emailAdd{zyz@itp.ac.cn}

\abstract{
Measuring the deviation from scale invariance of the primordial
power spectrum is a critical test of inflation.
In this paper we reconstruct the shape of the primordial power spectrum of
curvature perturbations from the cosmic microwave background
data, including the 7-year Wilkinson Microwave Anisotropy
Probe data and the Atacama Cosmology Telescope 148 GHz data,
by using a binning method of a cubic spline interpolation in
log-log space.
We find that the power-law spectrum is preferred by the data
and that the Harrison-Zel'dovich spectrum is disfavored at 95\%
confidence level. These conclusions hold with and without allowing for 
tensor modes, however the simpler model without tensors is preferred by the 
data. We do not find evidence for a feature in the primordial power 
spectrum -- in full agreement with generic predictions from cosmological 
inflation.}


\begin{document}
\maketitle

%%========================section 1 =========================
\section{Introduction}

Inflation in the early Universe predicts primordial power
spectra of curvature and tensor perturbations which lead to a
distinct imprint on the observed angular power spectrum of the
cosmic microwave background (CMB) radiation.
In the simplest models of slow-roll inflation the power spectra
can analytically be expressed in terms of the inflaton potential
and its first derivative at the Hubble scale crossing during inflation.
The angular power spectrum of the CMB is a convolution of the
primordial power spectrum with a radiative transfer function,
which can be obtained by integrating the full Einstein-Boltzmann
system of differential equations.
Therefore, given an inflaton potential or a primordial power spectrum,
the angular power spectrum of the CMB can numerically be calculated,
and thus the inflaton potential or the primordial power spectrum
can be fitted to CMB data~\cite{kom08,kom10,dun10}.

The simplest parameterization of the primordial power spectrum is
the phenomenological Harrison-Zel'dovich (scale invariant)
spectrum with one free parameter. The next simplest ansatz is to assume that there
is no distinguished scale, which leads to a simple power-law parameterization.
For this power-law ansatz, the scale-invariant spectrum is excluded at more
than 99\% confidence level by the Wilkinson Microwave Anisotropy Probe (WMAP)
data~\cite{kom10}.
Moreover, a slightly tilted power-law spectrum is still an
excellent fit to the data even adding a running index which
characterizes the deviation from a pure power law~\cite{kom10}.
Other specific parameterizations of the primordial power spectrum,
motivated by theoretical models or features of the observed data,
have been considered: for example, a broken power spectrum~\cite{bla03}
due perhaps to an interruption of the inflaton potential~\cite{bar01},
and a cutoff at large scales~\cite{efs03,bri06}
motivated by suppression of the lower multipoles in the CMB anisotropies~\cite{nol08,cop10}.

Since a modest deviation from scale invariance of the primordial power
spectrum is a critical prediction of inflationary models, it is a vital
test of the inflationary paradigm. Exact scale invariance would be expected for a
perfectly stationary epoch of inflation, which would never end. On the other hand, a
strong deviation from scale invariance could falsify the idea of cosmological inflation. Thus
it is important to probe the shape of the primordial power spectrum.
A strong theory prior on it could lead to misinterpretation and biases in
parameter determination.
Given our complete ignorance of the underlying physics of the very early Universe,
some model-independent approaches to reconstruct the shape
of the primordial power spectrum from existing data have been
employed, involving for example linear interpolation~\cite{bri03},
a minimally-parametric reconstruction~\cite{sea05},
wavelet expansions~\cite{muk03}, principle component analysis~\cite{hu03},
and a direct reconstruction via deconvolution methods~\cite{kog03,sha03,toc04}.
Clearly, multiple methods are needed to cross-check each other
and to contribute their respective strengths to our understanding.

In this work, we reconstruct the shape of the primordial power
spectrum using a binning method of a cubic spline interpolation
in the logarithmic amplitude and logarithmic wavenumber space,
where the power-law spectrum as a special case is just a
straight line.
This method guarantees the positivity of the power spectrum.
We use not only the seven-year WMAP data~\cite{kom10} but also
small-scale CMB data from the Atacama Cosmology Telescope
(ACT) experiment~\cite{dun10},
with two main astrophysical priors on the Hubble constant ($H_0$)
measured from the magnitude-redshift relation of 240 low-$z$
Type Ia supernovae at $z<0.1$~\cite{rie09} and on the distance ratios of
the comoving sound horizon to the angular diameter distances
from the Baryon Acoustic Oscillation (BAO) in the distribution
of galaxies~\cite{per09}.
Using data from ACT besides WMAP allows us to extend the measured
$k$ range to smaller scales. The combination of both data sets provides significant
improvements in the reconstruction of the spectrum.

In previous works, tensor modes have been commonly ignored when testing for the shape
of the primordial power spectrum. Here we consider the contribution from tensor modes and
assume a scale-invariant spectrum of tensor perturbations.
We compare different spectral shapes by means of the Akaike information criterion,
including a comparison between models with and without tensor perturbations. We show that
current data prefer a power-law shape and no tensor perturbations.

This paper is organized as follows. In Section~\ref{sec2},
we describe the method and the data used in this analysis.
In Section~\ref{sec3}, we present our results. Section~\ref{sec4} is devoted to
conclusions.

%%========================section 2 =========================
\section{Analysis method}\label{sec2}

We divide the primordial power spectrum of curvature perturbations,
$\PP_{\RR}$, into $N_{\rm bin}$ bins equally spaced in $\ln k$ between
$k_1=0.0002$ Mpc$^{-1}$ and $k_{N_{\rm bin}}=0.2$ Mpc$^{-1}$
for the 7-year WMAP data with $l \le 1200$ for the TT power spectrum
and $l \le 800$ for the TE power spectrum.
Moreover, we also use the low-$l$ temperature and polarization angular
power spectra from WMAP.
Adding the ACT data between $1000<l<3000$ allows a wide wavenumber
range with $k_{N_{\rm bin}}=0.3$ Mpc$^{-1}$.

To reconstruct a smooth spectrum with continuous first and second
derivatives with respect to $\ln k$, we use a cubic spline
interpolation to determine logarithmic values of the primordial
power spectrum for $\ln k_i < \ln k < \ln k_{i+1}$.
Here boundary conditions are adopted, where the second
derivative is set to zero.
For $k<k_1$ or $k>k_{N_{\rm bin}}$ we fix the slope of
the primordial power spectrum at the boundaries since the
CMB data place only weak constraints on them.
This reconstruction method has three advantages:
firstly, it is easy to detect deviations from a scale-invariant or
a power-law spectrum because both the scale-invariant and
power-law spectra are just straight lines
in the $\ln k$-$\ln \PP_{\RR}$ plane.
Secondly, negative values of the spectrum can be avoided by using $\ln \PP_{\RR}(k)$
instead of $\PP_{\RR}(k)$ for splines with steep slopes.
Thirdly, the shape of the power spectrum reduces to the scale-invariant or power-law
spectrum as a special case when $N_{\rm bin}=1,2$, respectively.
To summarize, we consider the following form of the power spectrum
in this work,
\bea
\ln \PP_{\RR}(k) = \left\{
\begin{array}{ll}
\left.\frac{d\ln \PP_{\RR}(k)}{d\ln k}\right|_{k_1} \ln \frac{k}{k_1} + \ln \PP_{\RR}(k_1), &  k<k_1; \\
\ln \PP_{\RR}(k_i), & k\in \{k_i\}; \\
{\rm cubic \ spline}, & k_i < k < k_{i+1};\\
\left.\frac{d\ln \PP_{\RR}(k)}{d\ln k}\right|_{k_{N_{\rm bin}}} \ln \frac{k}{k_{N_{\rm bin}}} + \ln \PP_{\RR}(k_{N_{\rm bin}}), & k>k_{N_{\rm bin}}.
\end{array}
\right.
\ena

We consider a spatially flat $\Lambda$CDM Universe described by
$N_{\rm bin}$ primordial spectrum parameters
$A_i \equiv \ln \left[10^{10} \PP_{\RR}(k_i)\right]$ and four background
parameters ($\Omega_b h^2$, $\Omega_c h^2$, $\Theta_s$, $\tau$), where
$\Omega_b h^2$ and $\Omega_c h^2$ are the physical baryon and cold
dark matter densities relative to the critical density,
$\Theta_s$ is the ratio of the sound horizon to the angular diameter
distance at decoupling, and $\tau$ is the reionization optical depth.
We also consider the Sunyaev-Zel'dovich (SZ) effect, in which CMB
photons scatter off hot electrons in clusters of galaxies.
Given a SZ template it is described by a SZ template amplitude
$A_{\rm SZ}$ as in the WMAP papers~\cite{kom08,kom10}.
For the 148 GHz ACT data~\cite{dun10}, aside from $A_{\rm SZ}$
we use, following the ACT analysis,
two more secondary parameters: $A_p$ and $A_c$.
The former is the total Poisson power at $l=3000$ from radio and
infrared point sources. The latter is the template amplitude of
the clustered power from infrared point sources.
If a contribution to the CMB from tensor modes is considered,
we assume a scale-invariant primordial power spectrum of tensor
perturbations. Current CMB measurements are insensitive to
deviations from this choice, as the tensor modes are subdominant in the TT angular
power spectrum and contribute at most up to $l \sim 100$.

Our analysis is carried out using a modified version of the
publicly available CosmoMC package, which explores the parameter space by
means of Monte Carlo Markov Chains~\cite{lew02}.
Besides the 7-year WMAP data ($l \le 1200$) including the low-$l$
temperature ($2 \le l \le 32$) and polarization ($2 \le l \le 23$)
data, we use two main astrophysical priors: the present-day
Hubble constant $H_0$ from the magnitude-redshift relation of
240 low-$z$ Type Ia supernovae at $z<0.1$~\cite{rie09}, and
the angular diameter distances out to $z=0.2$ and $0.35$,
measured from the two-degree field galaxy redshift survey and
the sloan digital sky survey data~\cite{per09}.
To enlarge the range in k, we use the WMAP data in combination
with the 148 GHz ACT data during its 2008 season.
For the ACT data, we focus on the band powers in the multiple
range $1000 \le l \le 3000$.
Following Ref.~\cite{dun10} for computational efficiency the
CMB is set to zero above $l=4000$ where the contribution is
subdominant, less than 5\% of the total power.
To use the 148 GHz ACT likelihood there are three secondary
parameters, $A_{\rm SZ}$, $A_p$ and $A_c$.
We impose positivity priors on these parameters and use
the SZ template and the clustered source template provided by
the ACT likelihood package.

To judge whether adding more bins effectively improves the fit,
we apply the Akaike information criterion (AIC) defined by
\bea
{\rm AIC} = -2\ln\LL_{\rm max}+2N_{\rm par},
\ena
where $\LL_{\rm max}$ is the maximum likelihood achievable by the model
and $N_{\rm par}$ the number of parameters.
This criterion arises from an approximation to the Kullback-Leibler
information entropy, used to measure the difference in goodness
of fit between two models. It sets up a tension between goodness of fit and
complexity of the model, with the best model minimizing the AIC.
Typically, models with too few parameters give a poor fit to
the data and hence have a low log-likelihood, while those with too
many are penalized by the second term.
Although it is not always clear how big a difference in AIC is
required for the worse model to be significantly disfavored,
typically a difference of 6 or more should be taken seriously.
This information criterion was used to carry out cosmological
model selection~\cite{lid04}.

%%========================section 3 =========================
\section{Results}\label{sec3}

\begin{figure}[!htb]
\begin{center}
\includegraphics[width=50mm]{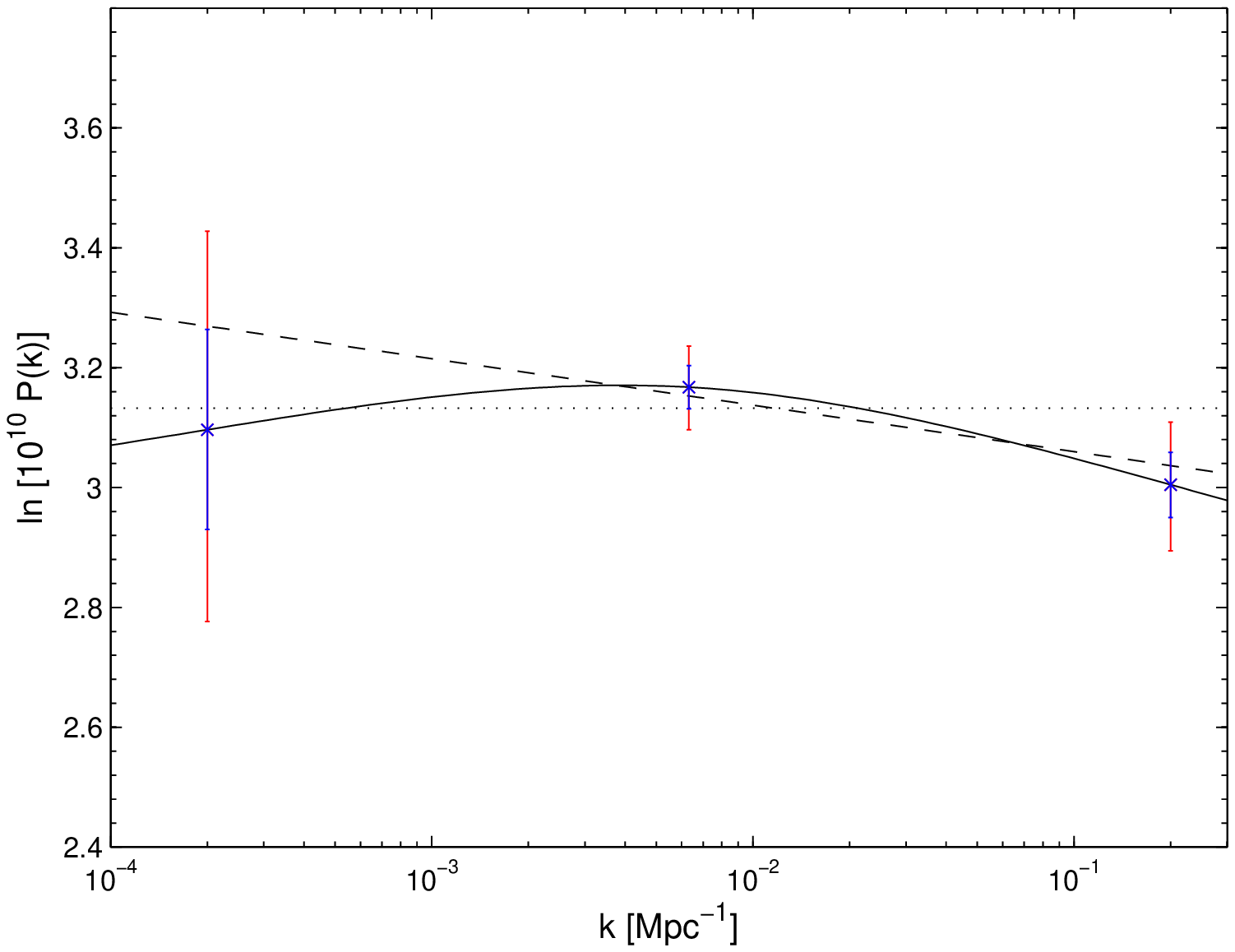}
\includegraphics[width=50mm]{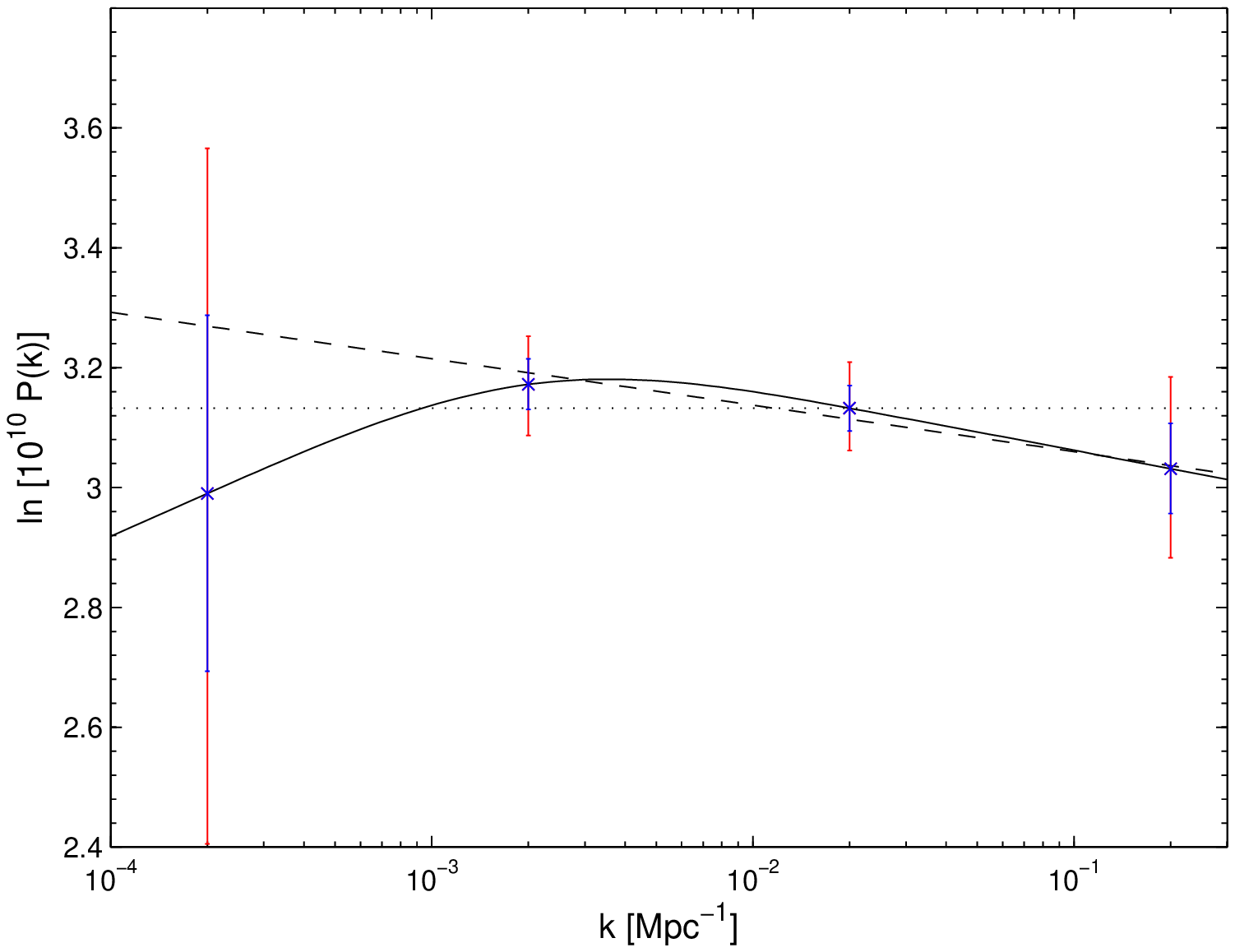}
\includegraphics[width=50mm]{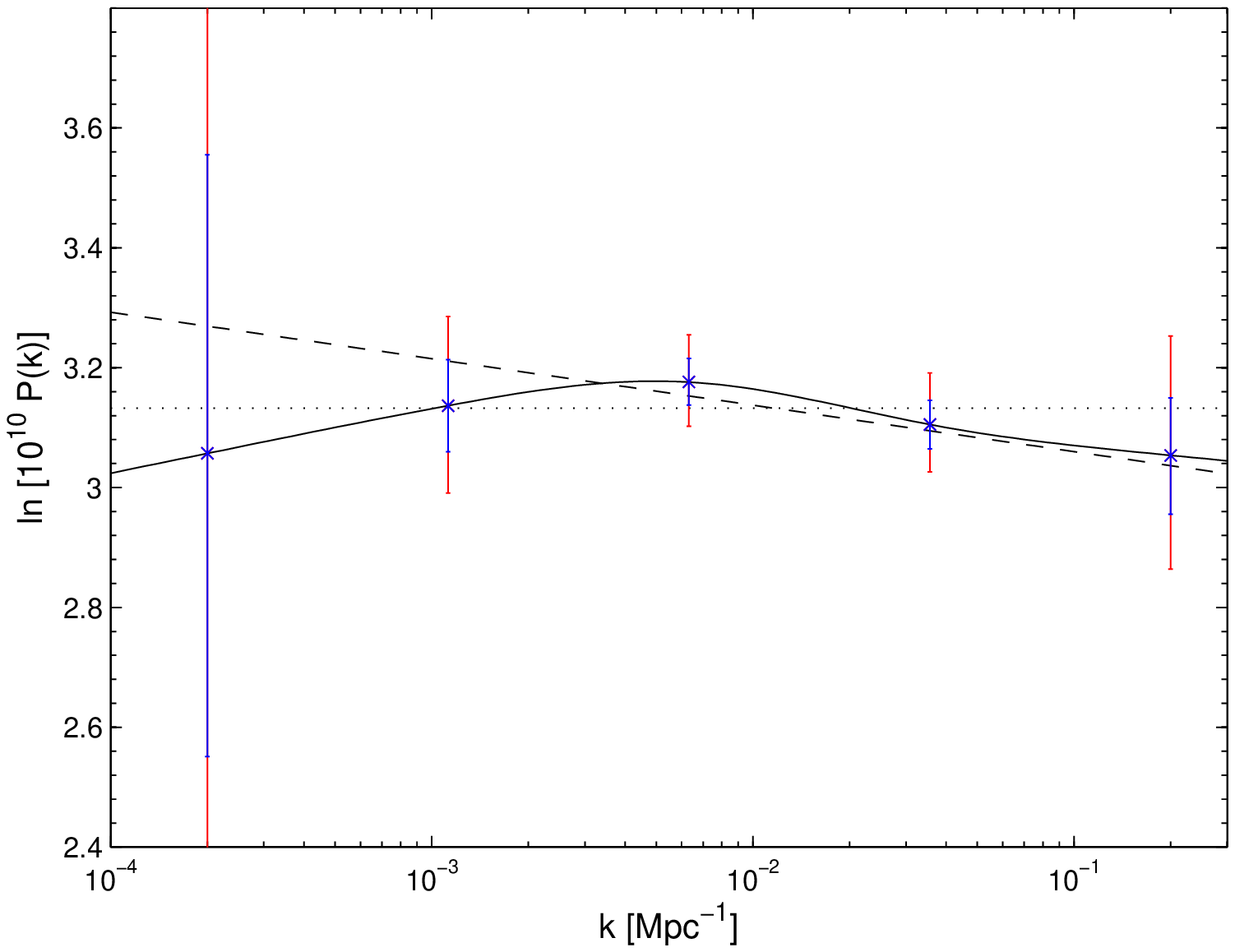}
\includegraphics[width=50mm]{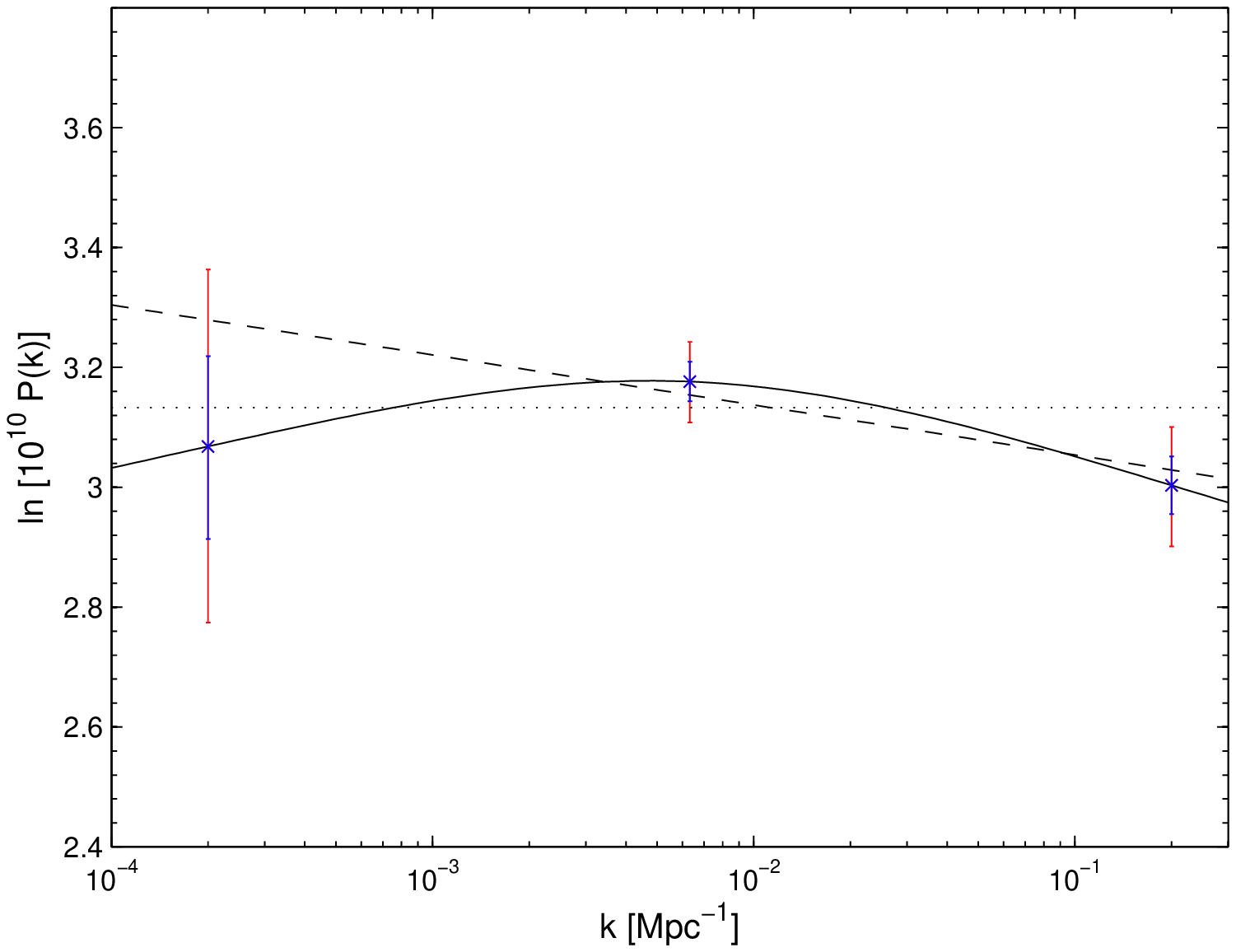}
\includegraphics[width=50mm]{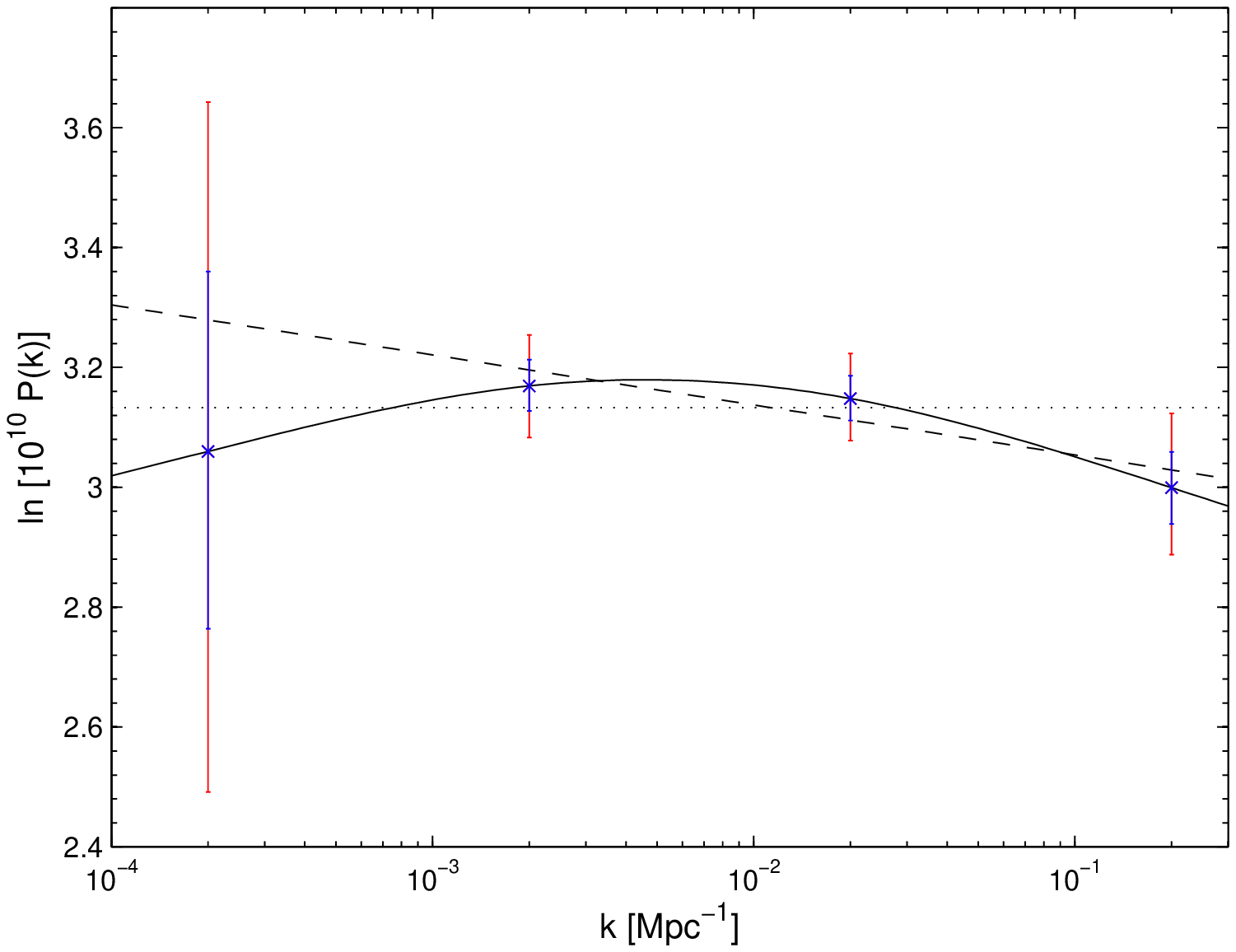}
\includegraphics[width=50mm]{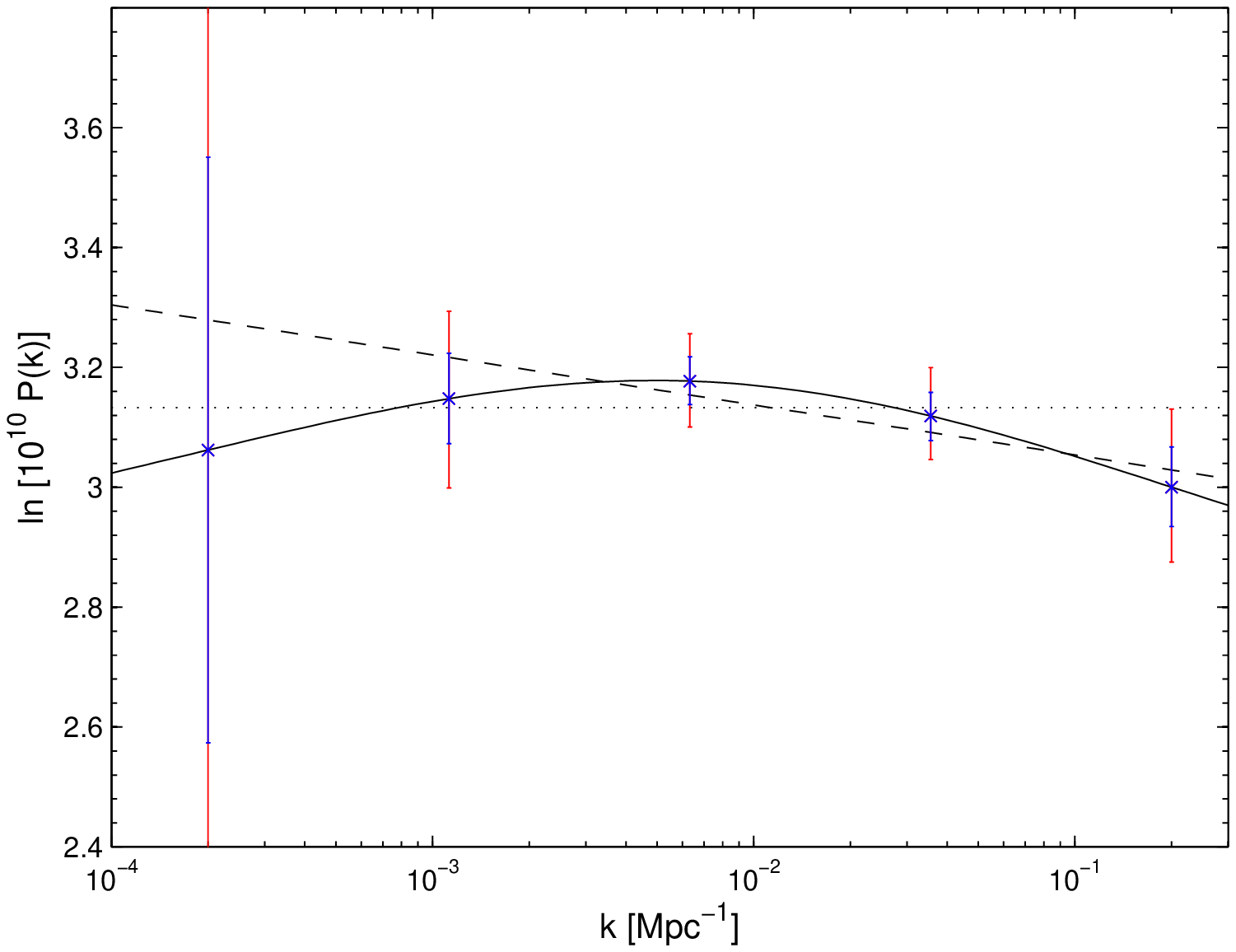}
\includegraphics[width=50mm]{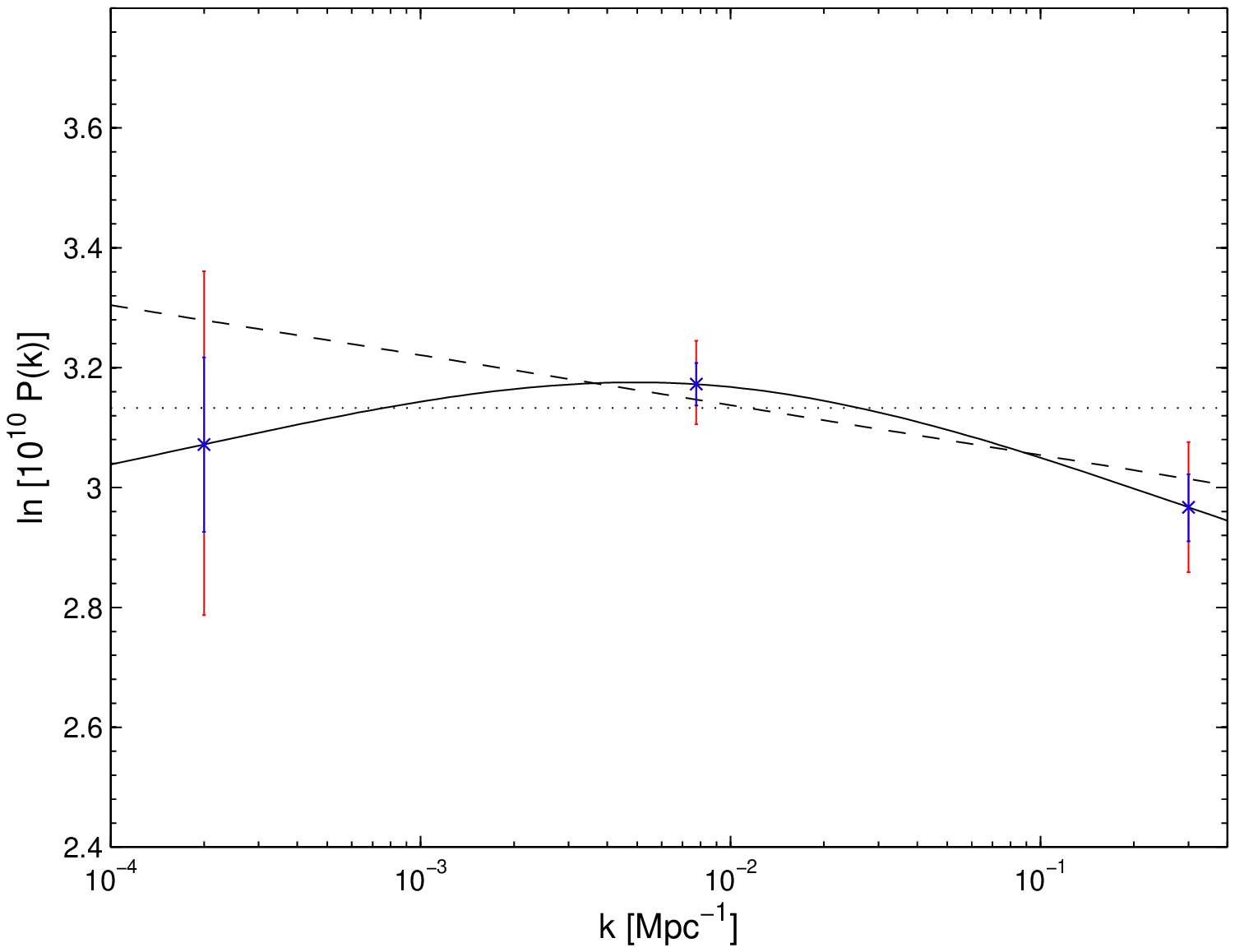}
\includegraphics[width=50mm]{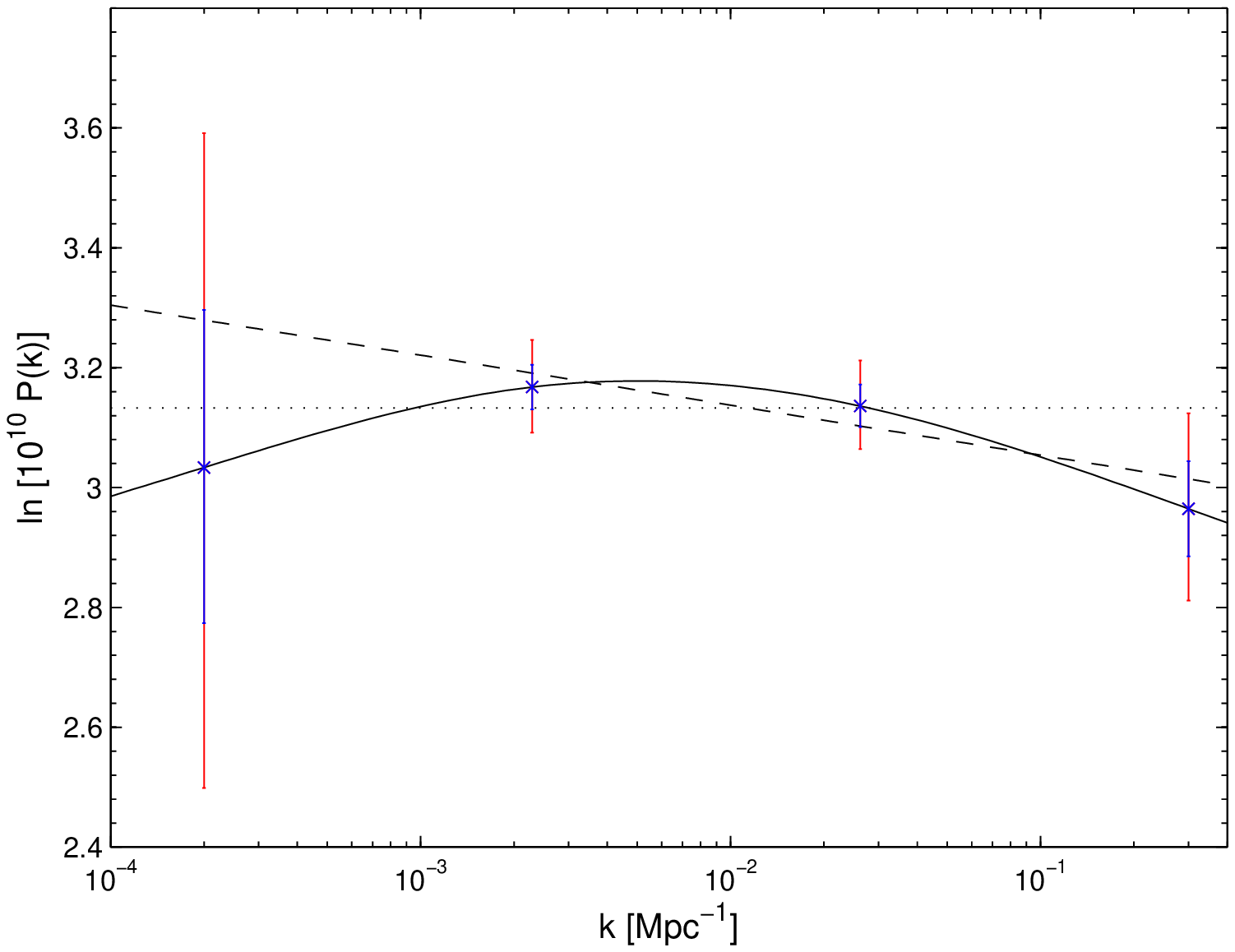}
\includegraphics[width=50mm]{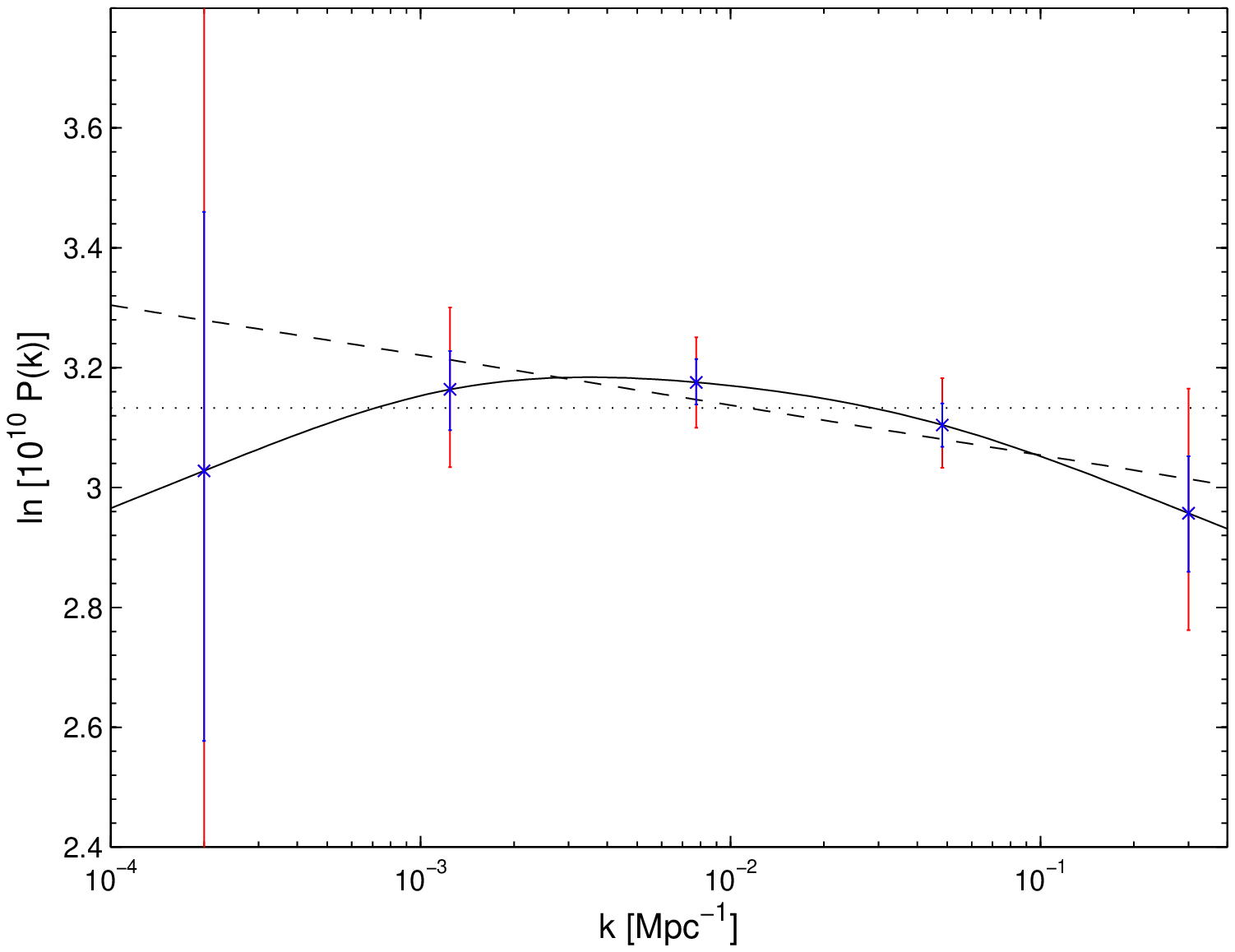}
\caption{Reconstruction of the primordial power spectrum without
tensor modes using a cubic spline interpolation for different binnings,
derived from the WMAP7+$H_0$+BAO combination (top panels) and
from the WMAP7+ACT+$H_0$+BAO combination for two different choices of 
$k_{N_{\rm bin}}$ (middle and bottom panels).
The dotted and dashed lines represent the scale-invariant ($N_{\rm bin} = 1$) and
power-law ($N_{\rm bin} = 2$) spectra respectively.
The blue and red error bars show $1\sigma$ and $2\sigma$ uncertainties
respectively. Note also that the best fit optical depth depends quite strongly on the assumed
shape of the primordial spectrum.}
\label{fig-scalar}
\end{center}
\end{figure}

Fig.~\ref{fig-scalar} shows the reconstructed power spectrum
of curvature perturbations without tensor modes, derived from
WMAP7+$H_0$+BAO (top panels) and WMAP7+ACT+$H_0$+BAO (middle and bottom panels).
Large uncertainties of the reconstructed spectrum at low-$l$
mainly arise from the cosmic variance while at high-$l$ different
systematics affect the reconstructed spectrum due to point source
emission and the SZ effect.
As the number of bins increases, the error bars of individual
bins grow due to the increase in degrees of freedom.
We can see that the power-law spectrum is an excellent fit to
the data while the scale-invariant spectrum is excluded by 95\%
confidence level at small scales if tensor modes are ignored.
Including the ACT data allows us to increase $k_{N_{\rm bin}}$ to
0.3 Mpc$^{-1}$.
At this scale the scale-invariant spectrum is far outside of
the 2$\sigma$ error bound.
As shown in Table~\ref{tab-scalar}, with the increasing number
of bins the AIC value first decreases and then increases for
$N_{\rm bin}>2$. We find that $N_{\rm bin} = 2$ (a power-law shape)
is the best model with and without ACT data and $N_{\rm bin} = 1$
(Harrison-Zel'dovich) provides the least favored fit to the data.
To demonstrate strong constraints on the spectrum at small scales from the ACT data,
we also reconstruct the spectrum in the case of $k_{N_{\rm bin}}=0.2$ Mpc$^{-2}$
in the middle panels of Fig.~\ref{fig-scalar}. Compared to the top panels, the errors in the last bin are reduced by including the ACT data.

\begin{table}[!htb]
\begin{center}
\begin{tabular}{lcccccc}
\hline
Data & $N_{\rm bin}$ & $N_{\rm par}$ & $-\ln \LL_{\rm max}$ & $\Delta \chi^2$ & AIC & $\Delta$AIC \\
\hline
WMAP7+$H_0$+BAO
 & 1 & 6  & 3742.4 & 0      & 7496.8 & 0      \\
 & 2 & 7  & 3738.7 & $-7.4$ & 7491.4 & $-5.4$ \\
 & 3 & 8  & 3738.4 & $-8.0$ & 7492.8 & $-4.0$ \\
 & 4 & 9  & 3738.4 & $-8.0$ & 7494.8 & $-2.0$ \\
 & 5 & 10 & 3738.4 & $-8.0$ & 7496.8 & 0 \\
\hline
+ACT
 & 1 & 8  & 3752.1 & 0       & 7520.2 & 0      \\
 & 2 & 9  & 3747.9 & $-8.4$  & 7513.8 & $-6.4$ \\
 & 3 & 10 & 3747.1 & $-10.0$ & 7514.2 & $-6.0$ \\
 & 4 & 11 & 3747.2 & $-9.8$  & 7516.4 & $-3.8$ \\
 & 5 & 12 & 3747.2 & $-9.8$  & 7518.4 & $-1.8$ \\
\hline
\end{tabular}
\end{center}
\caption{The $\chi^2$, AIC values and their differences with respect
to the scale-invariant power spectrum for various models,
derived from WMAP7+$H_0$+BAO and from WMAP7+ACT+$H_0$+BAO.
Here the contribution from tensor modes are ignored.}
\label{tab-scalar}
\end{table}

\begin{figure}[!htb]
\begin{center}
\includegraphics[width=50mm]{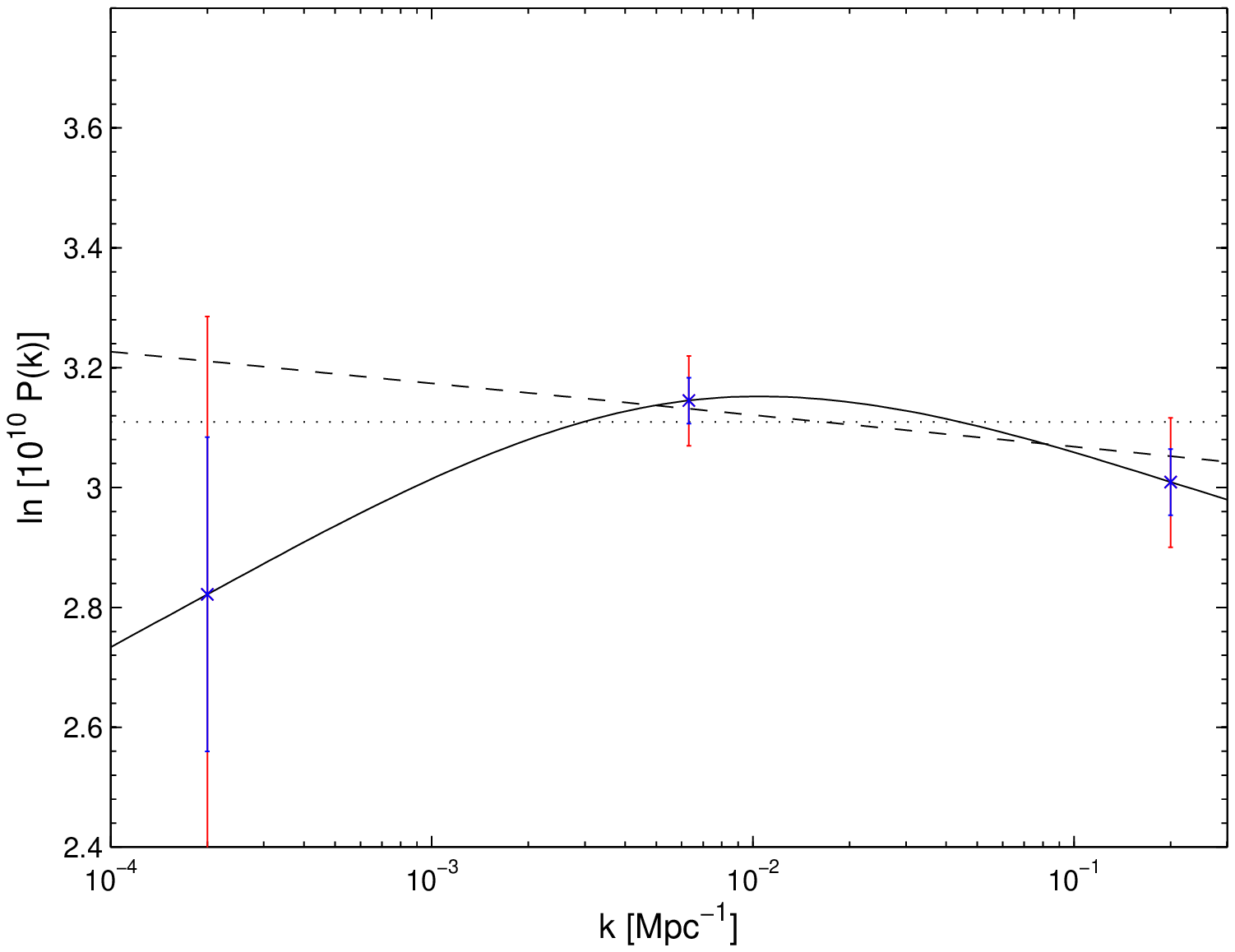}
\includegraphics[width=50mm]{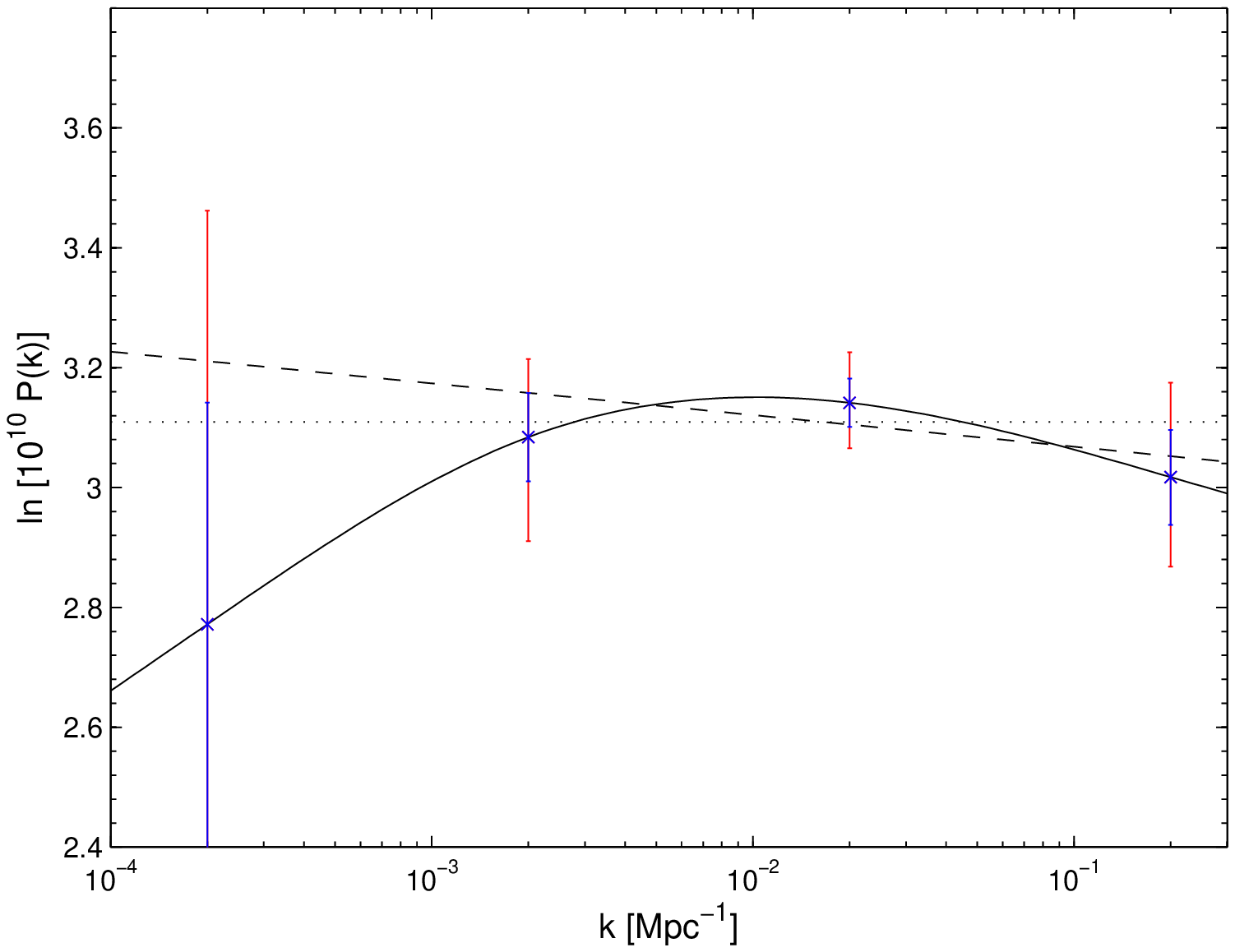}
\includegraphics[width=50mm]{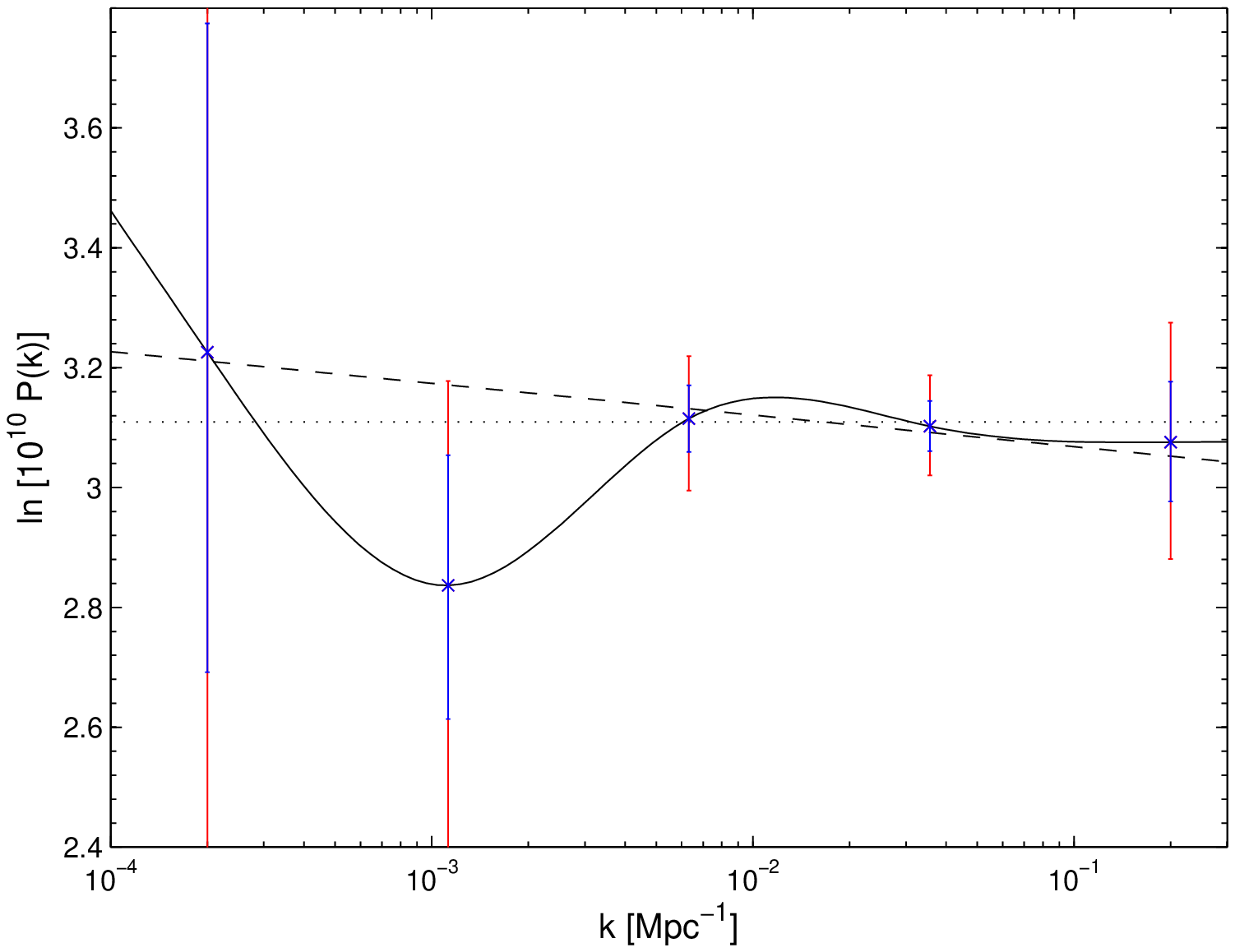}
\includegraphics[width=50mm]{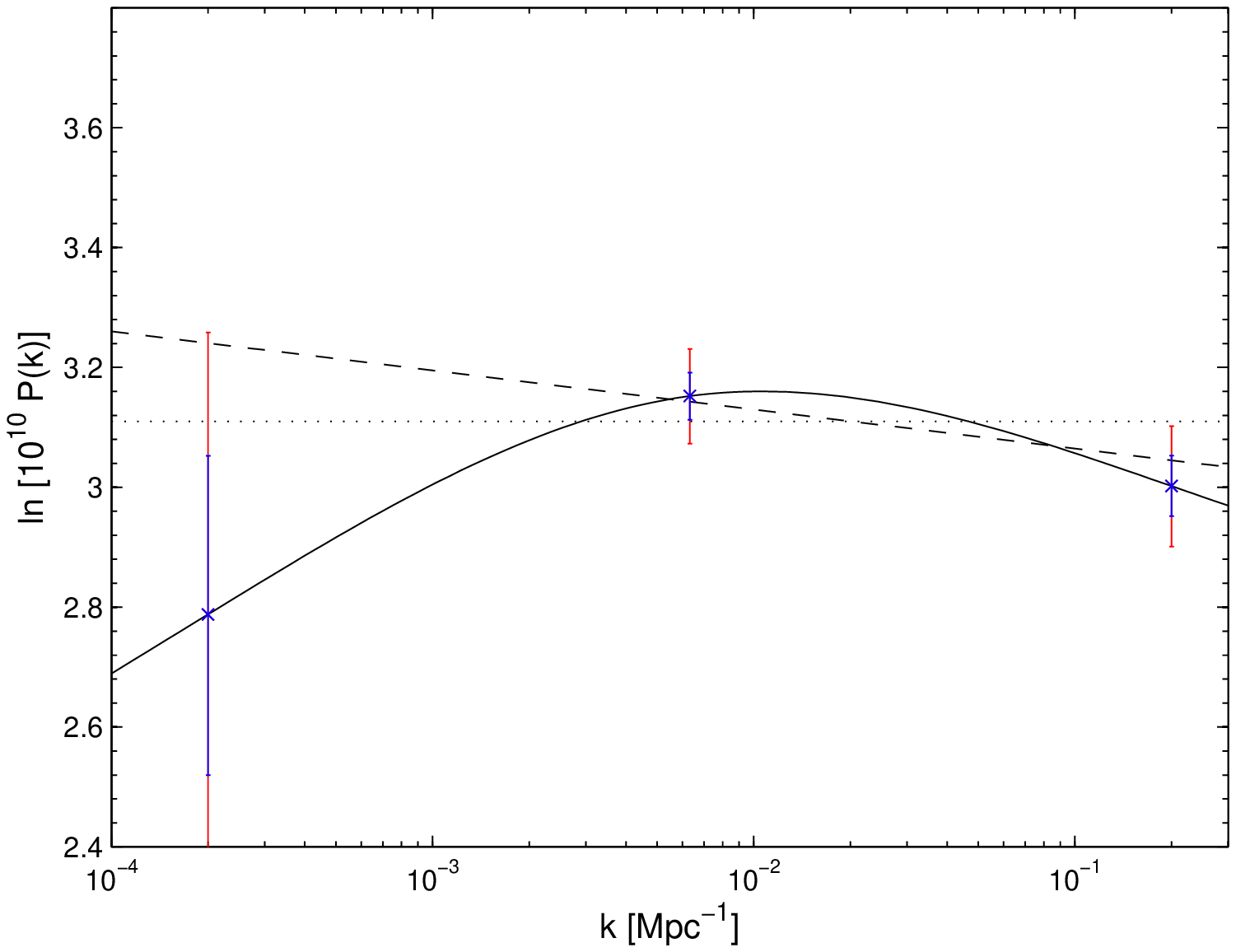}
\includegraphics[width=50mm]{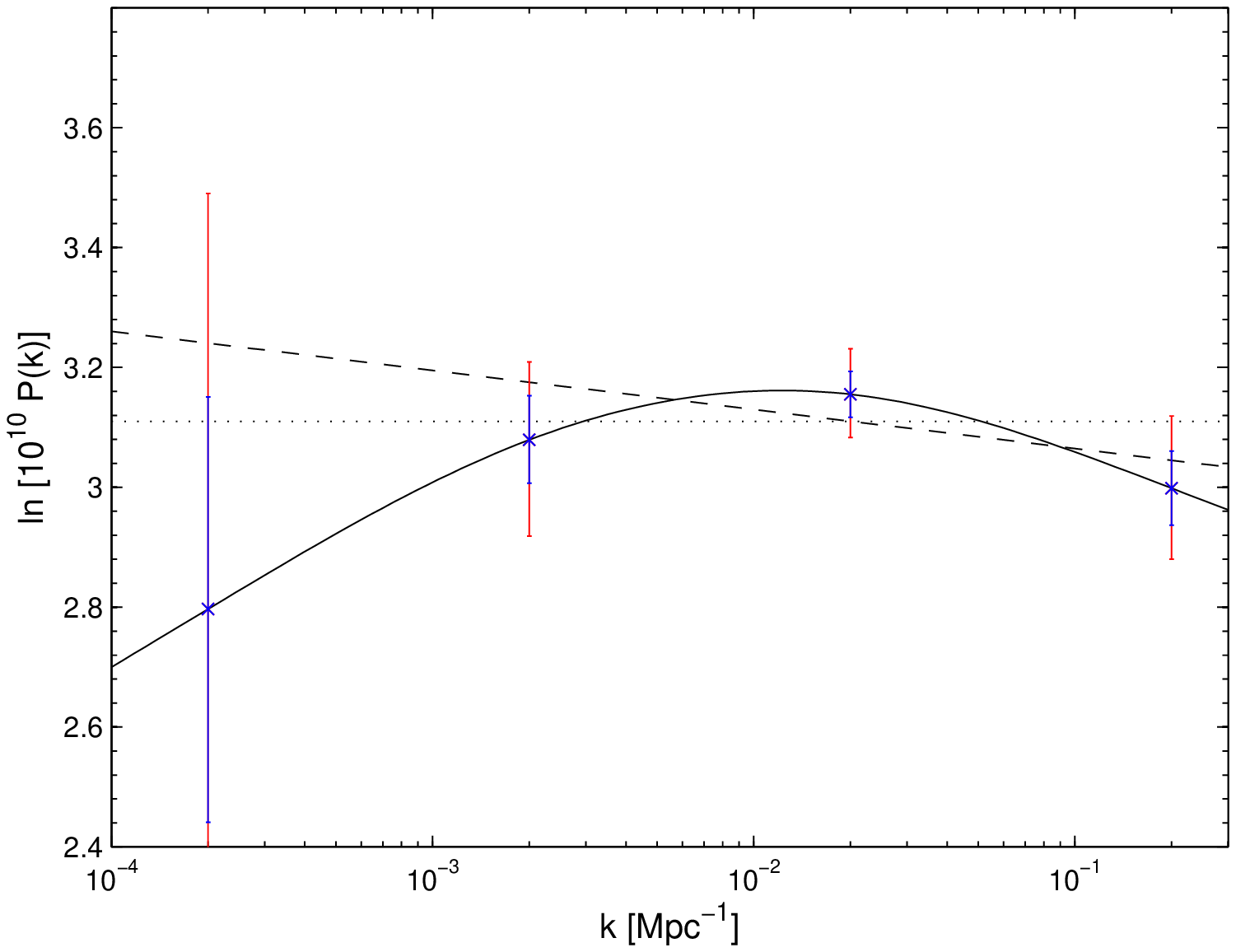}
\includegraphics[width=50mm]{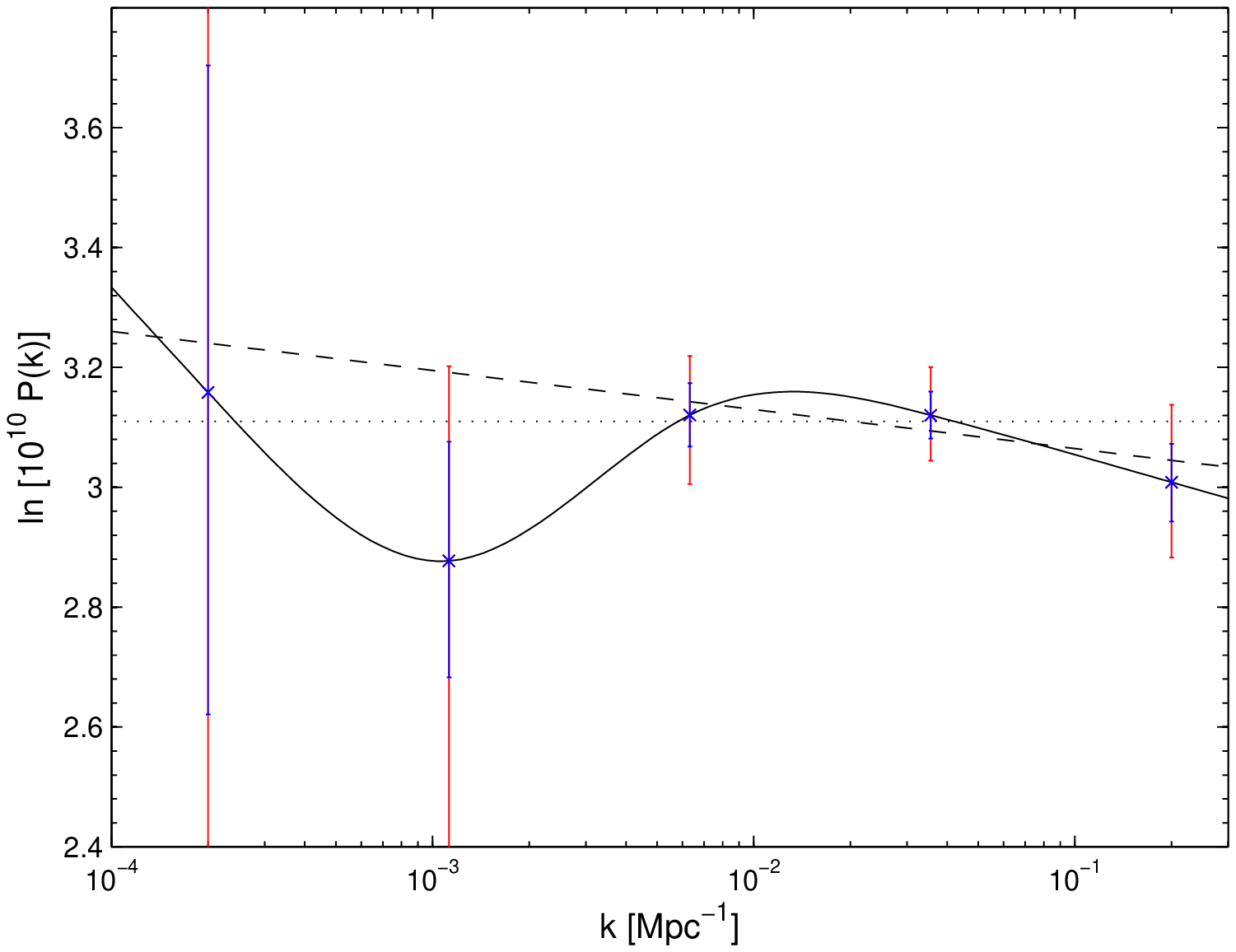}
\includegraphics[width=50mm]{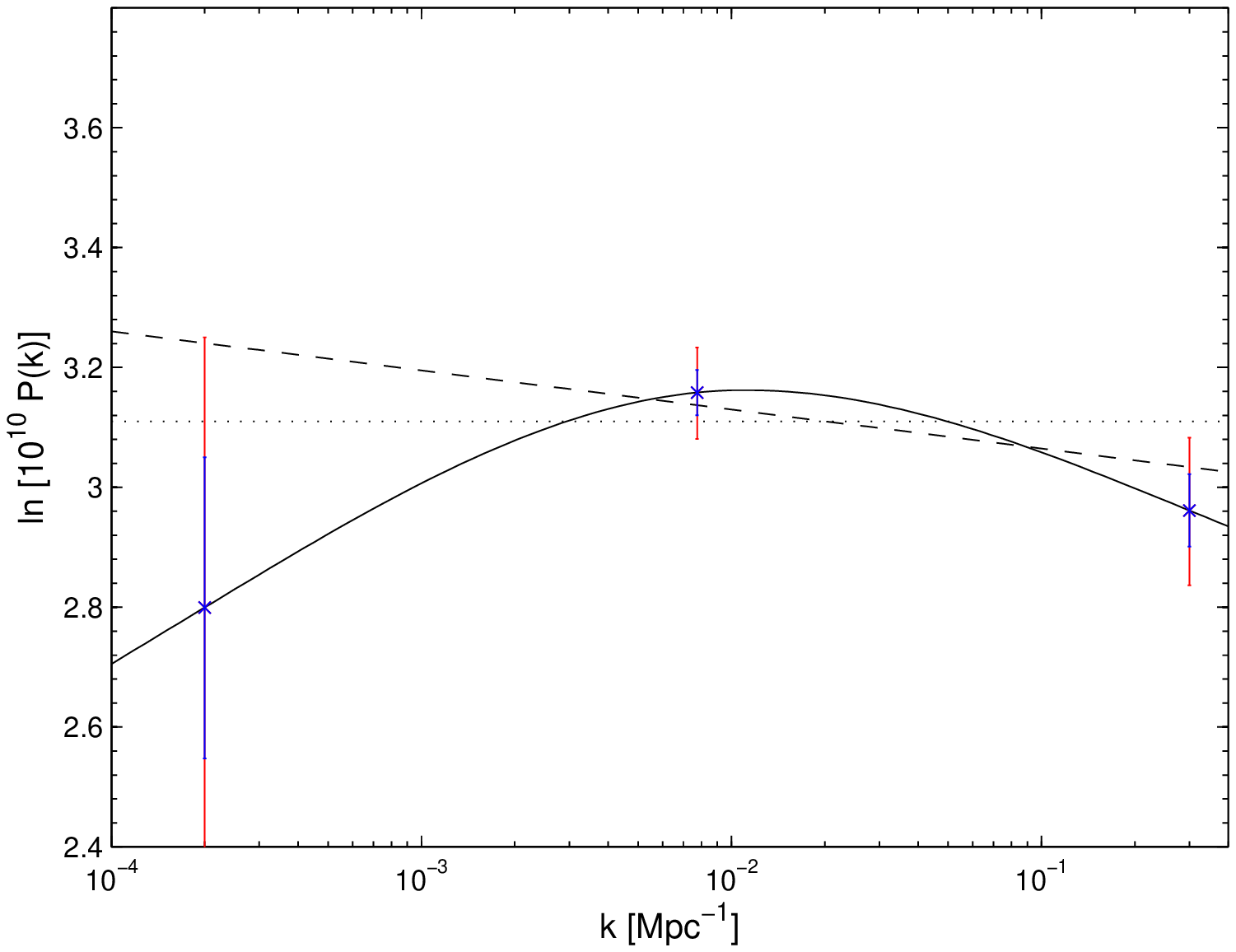}
\includegraphics[width=50mm]{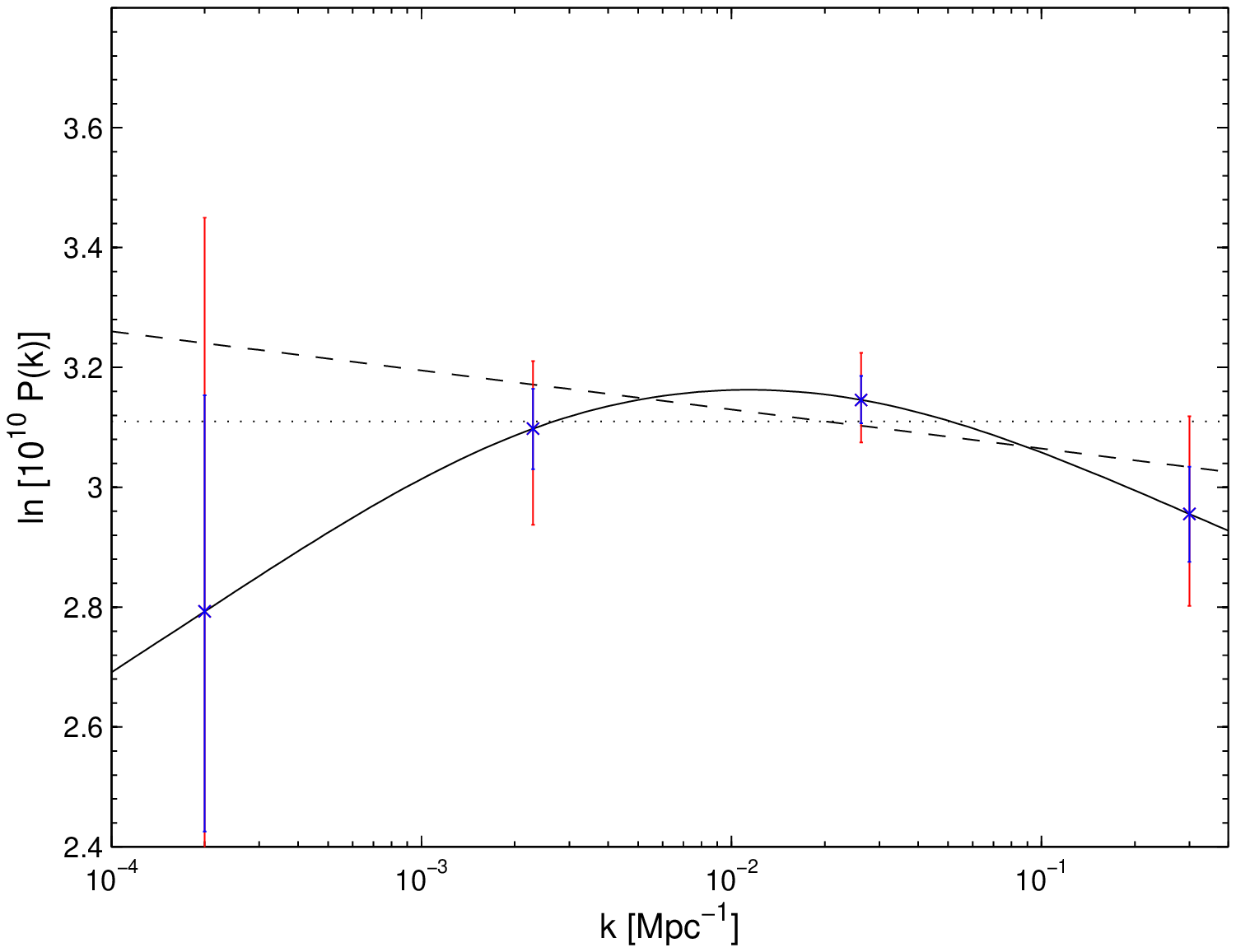}
\includegraphics[width=50mm]{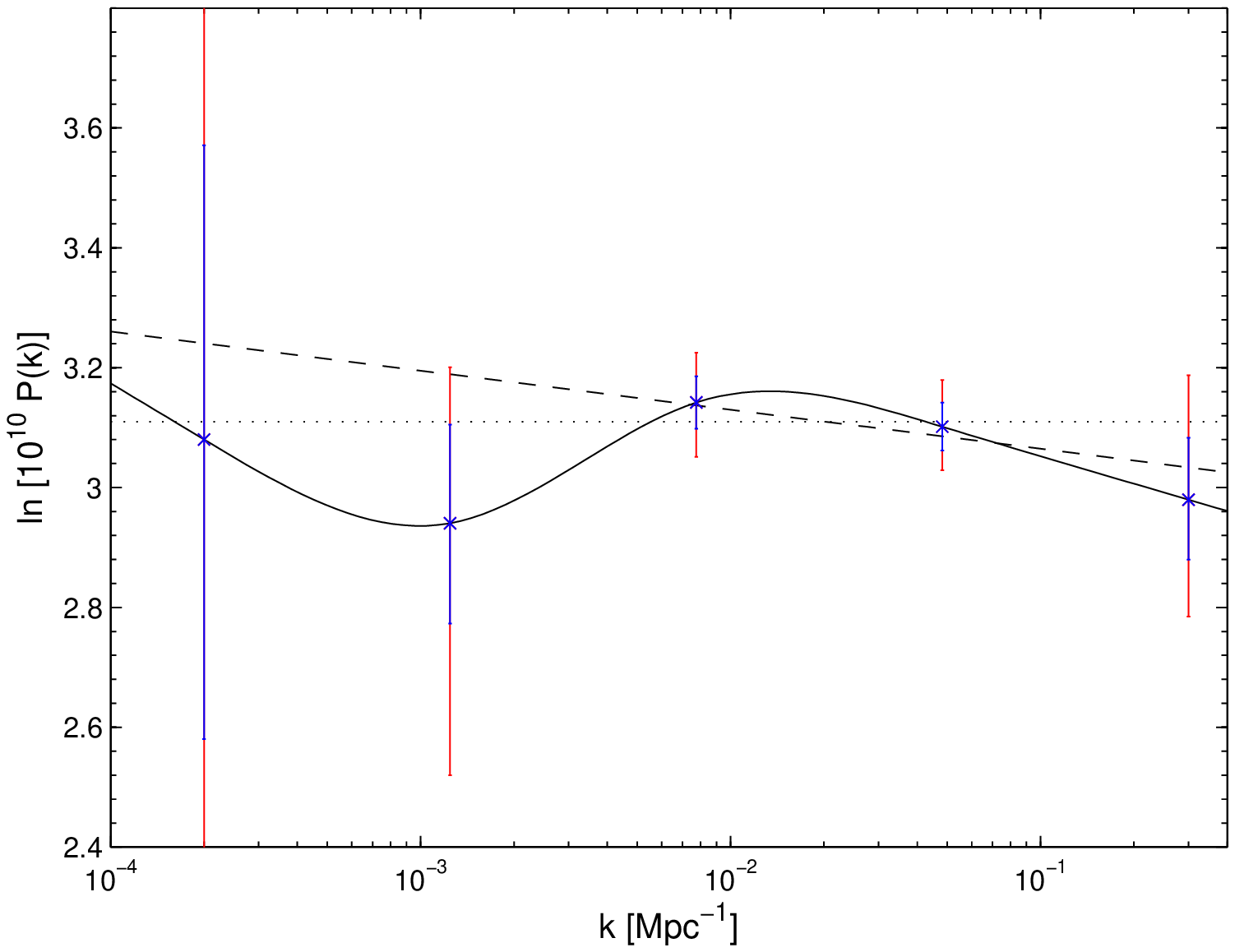}
\caption{Reconstruction of the primordial curvature power spectrum with
a scale-invariant spectrum of tensor perturbations using a
cubic spline interpolation for different binnings,
derived from WMAP7+$H_0$+BAO (top panels) and
from WMAP7+ACT+$H_0$+BAO for two different choices of 
$k_{N_{\rm bin}}$ (middle and bottom panels).
The dotted and dashed lines represent the scale-invariant and
power-law spectra respectively.
The blue and red error bars show $1\sigma$ and $2\sigma$ uncertainties
respectively.}
\label{fig-tensor}
\end{center}
\end{figure}

Assuming a scale-invariant spectrum of tensor perturbations,
the reconstructed primordial power spectrum of curvature perturbations
are shown in Fig.~\ref{fig-tensor}.
Due to the contribution of tensor modes to the angular power spectra
of the CMB, the amplitude of the Harrison-Zel'dovich spectrum is
smaller than one in Fig.~\ref{fig-scalar}.
For WMAP7+$H_0$+BAO, there is no convincing deviation from either a
simple scale-invariant or a power-law spectrum as shown in the top
panels of Fig.~\ref{fig-tensor}.
As expected, including the ACT data will improve the measurements
of the spectrum.
In the case of $N_{\rm bin}=3$ the reconstructed spectrum deviates
from the scale-invariant one on small scales at 95\% confidence level.
In the $N_{\rm bin}=4$ case, since the error bars become larger,
it becomes consistent with the scale-invariant spectrum.
From Table~\ref{tab-tensor}, we can see that increasing the number
of bins beyond 2 increases the AIC value slightly.
Hence, under the assumption of a scale-invariant tensor spectrum,
the power-law spectrum provides a better fit to the data.
Comparing Table~\ref{tab-tensor} with Table~\ref{fig-scalar},
we find that the scale-invariant spectrum with tensor modes is more
favored than one without tensor modes while the power-law spectrum
without tensor modes is more favored than one with tensor modes.

\begin{table}[!htb]
\begin{center}
\begin{tabular}{lccccccc}
\hline
Data & $N_{\rm bin}$ & $N_{\rm par}$ & $-\ln \LL_{\rm max}$ & $\Delta \chi^2$ & AIC & $\Delta$AIC \\
\hline
WMAP7+$H_0$+BAO
 & 1 & 7  & 3740.5 & $-3.8$ & 7495.0 & $-1.8$ \\
 & 2 & 8  & 3738.8 & $-7.2$ & 7493.6 & $-3.2$ \\
 & 3 & 9  & 3738.4 & $-8.0$ & 7494.8 & $-2.0$ \\
 & 4 & 10 & 3738.4 & $-8.0$ & 7496.8 & $0$ \\
 & 5 & 11 & 3738.2 & $-8.4$ & 7498.4 & $1.6$ \\
\hline
+ACT & 1 & 9  & 3750.4 & $-3.4$  & 7518.8 & $-1.4$ \\
     & 2 & 10 & 3747.9 & $-8.4$  & 7515.8 & $-4.4$ \\
     & 3 & 11 & 3747.1 & $-10.0$ & 7516.2 & $-4.0$ \\
     & 4 & 12 & 3747.1 & $-10.0$ & 7518.2 & $-2.0$ \\
     & 5 & 13 & 3747.2 & $-9.8$  & 7520.4 & $0.2$ \\
\hline
\end{tabular}
\end{center}
\caption{The $\chi^2$, AIC values and their differences with respect
to the scale-invariant power spectrum without tensor modes for various models with tensor modes,
derived from WMAP7+$H_0$+BAO and from WMAP7+ACT+$H_0$+BAO.
We assume a scale-invariant spectrum of tensor perturbations.}
\label{tab-tensor}
\end{table}

In Table~\ref{tab-params} we summarize the estimated primordial
power spectrum values as well as the other cosmological parameters.
Note that the best fit reionization optical depth depends quite
strongly on the assumed shape of the primordial spectrum.
As $N_{\rm bin}$ increases from 1 to 3, the reionization optical
depth first decreases and then increases drastically.

\begin{table}[!htb]
\scriptsize
\begin{center}
\begin{tabular}{lccccccccc}
\hline
$N_{\rm bin}$ & $A_1$ & $A_2$ & $A_3$ & $A_4$ & $A_5$ & $\O_bh^2$ & $\O_ch^2$ & $\O_\Lambda$ & $\tau$ \\
\hline
\multicolumn{4}{l}{WMAP7+$H_0$+BAO  ($k_{N_{\rm bin}}=0.2$ Mpc$^{-1}$)} \\
 1  & 3.132$\pm$0.031 & -- & -- & -- & -- & 0.02374 & 0.1131 & 0.738 & 0.101 \\
 2  & 3.269$\pm$0.060 & 3.037$\pm$0.045 & -- & -- & -- & 0.02261 & 0.1125 & 0.727 &  0.088 \\
 3  & 3.096$\pm$0.167 & 3.168$\pm$0.032 & 3.005$\pm$0.054 & -- & -- & 0.02224 & 0.1137 & 0.719 & 0.092 \\
 4  & 2.990$\pm$0.296 & 3.172$\pm$0.042 & 3.132$\pm$0.038 & 3.031$\pm$0.076 & -- & 0.02233 & 0.1130 & 0.722 & 0.094 \\
 5  & 3.057$\pm$0.505 & 3.137$\pm$0.076 & 3.176$\pm$0.039 & 3.105$\pm$0.042 & 3.054$\pm$0.099 & 0.02224 & 0.1127 & 0.722 & 0.095 \\
\multicolumn{4}{l}{+Tensor}\\
 1  & 3.109$\pm$0.033 & -- & -- & -- & -- & 0.02364 & 0.1117 & 0.741 & 0.094 \\
 2  & 3.211$\pm$0.074 & 3.052$\pm$0.048 & -- & -- & -- & 0.02289 & 0.1117 & 0.733 & 0.087 \\
 3  & 2.822$\pm$0.267 & 3.145$\pm$0.038 & 3.009$\pm$0.055 & -- & -- & 0.02244 & 0.1137 & 0.722 & 0.097 \\
 4  & 2.772$\pm$0.380 & 3.084$\pm$0.077 & 3.141$\pm$0.041 & 3.017$\pm$0.079 & -- &  0.02251 & 0.1132 & 0.724 & 0.098 \\
 5  & 3.226$\pm$0.549 & 2.837$\pm$0.227 & 3.115$\pm$0.057 & 3.102$\pm$0.042 & 3.076$\pm$0.100 & 0.02219 & 0.1124 & 0.724 & 0.095 \\
\hline
\multicolumn{4}{l}{WMAP7+ACT+$H_0$+BAO ($k_{N_{\rm bin}}=0.3$ Mpc$^{-1}$)} \\
 1  & 3.133$\pm$0.032 & -- & -- & -- & -- & 0.02358 & 0.1124 & 0.741 & 0.102 \\
 2  & 3.279$\pm$0.056 & 3.014$\pm$0.047 & -- & -- & -- & 0.02239 & 0.1125 & 0.727 & 0.085 \\
 3  & 3.072$\pm$0.148 & 3.173$\pm$0.036 & 2.967$\pm$0.056 & -- & -- & 0.02203 & 0.1143 & 0.716 & 0.094 \\
 4  & 3.033$\pm$0.274 & 3.168$\pm$0.039 & 3.136$\pm$0.037 & 2.965$\pm$0.080 & -- & 0.02204 & 0.1146 & 0.715 & 0.094 \\
 5  & 3.028$\pm$0.452 & 3.164$\pm$0.068 & 3.176$\pm$0.039 & 3.104$\pm$0.038 & 2.957$\pm$0.099 & 0.02210 & 0.1148 & 0.714 & 0.095 \\
\multicolumn{4}{l}{+Tensor}\\
 1  & 3.110$\pm$0.031 & -- & -- & -- & -- & 0.02347 & 0.1109 & 0.745 & 0.095 \\
 2  & 3.241$\pm$0.048 & 3.034$\pm$0.037 & -- & -- & -- & 0.02253 & 0.1117 & 0.731 & 0.086 \\
 3  & 2.800$\pm$0.258 & 3.158$\pm$0.039 & 2.961$\pm$0.062 & -- & -- & 0.02225 & 0.1146 & 0.718 & 0.098 \\
 4  & 2.793$\pm$0.374 & 3.098$\pm$0.069 & 3.146$\pm$0.039 & 2.956$\pm$0.080 & -- & 0.02227 & 0.1148 & 0.717 & 0.098 \\
 5  & 3.080$\pm$0.498 & 2.941$\pm$0.173 & 3.142$\pm$0.044 & 3.102$\pm$0.039 & 2.980$\pm$0.102 & 0.02204 & 0.1152 & 0.713 & 0.094 \\
\hline
\end{tabular}
\end{center}
\caption{Mean values and marginalized 68\% confidence levels
for the primordial spectrum parameters, and the mean values for
the background parameters.}
\label{tab-params}
\end{table}

%%========================section 4 =========================
\section{Conclusions}\label{sec4}

In this work we have reconstructed the smooth shape of the
primordial power spectrum of curvature perturbations without
or with tensor modes from the 7-year WMAP data in combination
with the small-scale CMB data from the ACT experiment.
We adopt the cubic spline interpolation method in the
$\ln k$-$\ln \PP_\RR$ plane to avoid negative values of the
amplitude and to make the detection of deviations from the power law easy.
If we ignore tensor modes, the scale-invariant spectrum is
excluded by the combined data at 95\% confidence level,
which is consistent with the result of Ref.~\cite{kom10}.
Moreover, we find no convincing deviation from a simple power-law
spectrum as suggested in Refs.~\cite{sea05,sha03}.
If tensor modes are included, for WMAP7+$H_0$+BAO the
scale-invariant spectrum lies within the 2$\sigma$ bound.
For WMAP7+ACT+$H_0$+BAO,
there a deviation from the scale-invariant spectrum
at small scales is found at the 95\% confidence level under the assumption
of a scale-invariant tensor spectrum.

Comparing Fig.~\ref{fig-scalar} with Fig.~\ref{fig-tensor},
we find that including tensor modes enhances the bending
of the primordial power spectrum.
Moreover, the AIC values listed in Table~\ref{tab-scalar} and
Table~\ref{tab-tensor} indicate that the power-law spectrum
without tensor modes is the best fit to the combined data.
We do not find evidence for a feature in the primordial power spectra,
although a bent spectrum with a suppression of low $k$ and
high $k$ modes with respect to a power-law spectrum is only slightly disfavored
by the AIC ($-6.4$ vs.~$-6.0$ without tensors and $-4.4$ vs.~$-4.0$ with tensors).
Without ACT the case for a bend is weaker.

We emphasize that our reconstruction method is insensitive
to local features in the primordial power spectrum,
but is sensitive to the overall shape, since the cubic spline
interpolation makes the primordial power spectrum smoother
than linear interpolation~\cite{bri03}.
Therefore, it is complementary to wavelet expansions~\cite{muk03}
and principle component analysis~\cite{hu03}, which provide
rigorous methods to search for sharp features.
This type of analysis is also complementary to the direct testing
of slow-roll inflation~\cite{mar04,haz10} and the analysis of
the WMAP team~\cite{kom08,kom10}.

\acknowledgments
We thank E.~Komatsu and J.~Dunkley for useful discussions.
%Our numerical analysis was performed on the HPC cluster of the RWTH Aachen.
This work was supported in part by the Alexander von Humboldt Foundation.
ZKG is partially supported by the project of Knowledge Innovation
Program of Chinese Academy of Science and
National Basic Research Program of China under Grant No:2010CB832805. DJS
acknowledges support by Deutsche Forschungsgemeinschaft (DFG).
We used CosmoMC and CAMB.
We also acknowledge the use of WMAP data from the LAMBDA server and ACT data.

\paragraph{Note Added:}
When this paper was concluded, we became aware that the ACT and WMAP data
were used to reconstruct the primordial power spectrum in~\cite{hlo11}.

\end{document}